\documentclass[12pt,aps,prb,preprint,showpacs,showkeys]{revtex4}

\usepackage{graphicx}
\begin{document}
\draft
\title{New magnetostatic modes in small nonellipsoidal magnetic particles}
\author{H. Puszkarski\cite{correspond}, M. Krawczyk}
\address{Surface Physics Division, Faculty of Physics, Adam
Mickiewicz University, \\ ul. Umultowska 85, Pozna\'{n}, 61-614 Poland.}
\author{J.-C. S. L\'{e}vy}
\address{Laboratoire "Mat\'{e}riaux et Ph\'{e}nom\`{e}nes Quantiques" (MPQ) , case 7021, Universit\'{e} Paris 7, 2 place Jussieu, 75251 Paris
C\'{e}dex 05 France.}

\date{\today}

\begin{abstract}
Magnetostatic normal modes are investigated here in elongated rods. The dipolar field resulting from the dipole-dipole interactions is
calculated numerically in points of the axis connecting opposite rod face centers (\emph{central axis}) by collecting individual contributions
to this field coming from each of the atomic planes perpendicular to the central axis. The applied magnetic field is assumed to be oriented
along the central axis, and the magnetization to be uniform throughout the sample. The \emph{frequency} spectrum of magnetostatic waves
propagating in the direction of the applied field is found numerically by solving the Landau-Lifshith equation of motion with the spatially
\emph{nonhomogeneous} dipolar field taken into account; the mode amplitude \emph{profiles} are depicted as well. While energetically highest
modes have \emph{bulk-extended} character, the modes forming the lower part of the spectrum are localized in the subsurface region
(\emph{bulk-dead modes}). Between these two mode types, magnetostatic modes of a new type (\emph{comb modes}) are found to occur, characterized
by two clearly discernible regions: a zone of fast amplitude oscillations inside the rod, and narrow slow-oscillation regions at the borders.
Absorbing virtually no energy from an applied alternating field, comb modes will have no significant contribution to the magnetic noise.

\end{abstract}
\keywords{magnetostatic modes; magnetic nanograins; mode localization; demagnetizing field}  \pacs{75.30.Ds, 75.40.Gb, 75.75.+a} \maketitle
%% maketitle must follow the abstract.

%% If there is not enough space inside the running head
%% for all authors including the title you may provide
%% the leftmark in one of the following three forms:

%% \renewcommand{\leftmark}
%% {First Author: A Short Title}

%% \renewcommand{\leftmark}
%% {First Author and Second Author: A Short Title}

\section{Introduction}

Modern magnetic storage media require the use of very small magnetic elements (see {\em e.g.} [\onlinecite{givord03}]). When the recording
density approaches 100 $Gb/in^{2}$ the bit size is around $0.2\mu m \times 0.1\mu m$, and similar size is required for the read heads. On the
other hand, when the bit size becomes smaller, the sensitivity of the read heads should increase in order to detect the weaker magnetic field
produced by the bit. In small nonellipsoidal magnetic particles the internal field nonhomogeneity, due to the particle shape and produced by
long-range dipolar interactions, has to be taken into account. It is known that dipolar size and shape effect present in the magnetostatic wave
spectra of such particles result in additional power losses \cite{2} very important in device applications. Therefore, one way to improve the
quality of the read heads is to suppress their magnetic noise. In this paper we show that the magnetostatic mode spectrum of the elongated
nanorods reveal specific range of frequencies when new type of excitations (we shall call them "comb modes") are accessible and when they are
excited by the AC magnetic field the absorbed power is negligible. Therefore,  in this particular range of frequencies, magnetostatic modes will
not contribute to the level of the magnetic noise.

    The dynamical properties of small magnetic elements have been actively investigated recently
    in studies using different methods and approaches (see our recent papers [\onlinecite{levy04, puszkarski05}] and references therein).
    These different methods give good indications of the localized magnetostatic mode existence. This paper is focused on
    localization properties of pure magnetostatic modes in elongated rods (having a nanometric square cross-section).
    It should be emphasized that the localization effects we are going to discuss here can be revealed only in magnetic systems in
    which dipolar interactions are dominant, and for this reason, exchange interactions are omitted here from the very beginning. If exchange
    interactions were taken into account in our investigations, the mode localization properties in question
    would disappear completely [\onlinecite{hp}].

    In our present theoretical approach the studied dipolar system is
    regarded as a set of discrete magnetic dipoles regularly arranged in a crystalline lattice.
    The dipolar energy is calculated by collecting contributions from each \emph{dipolar lattice plane} parallel
    to the rod base (see Fig. \ref{konfprost}). As the summing is performed over \emph{all} the dipoles within
    a given dipolar plane, no approximation is involved in our dipolar energy evaluation. The scheme of
    our calculations has already been presented in our former paper [\onlinecite{puszkarski05}], restricted to cubic systems.

\section{A finite system of magnetic dipoles in planar
arrangement}\label{section1}

\begin{figure}
%\begin{center}
\includegraphics[width=7cm ]{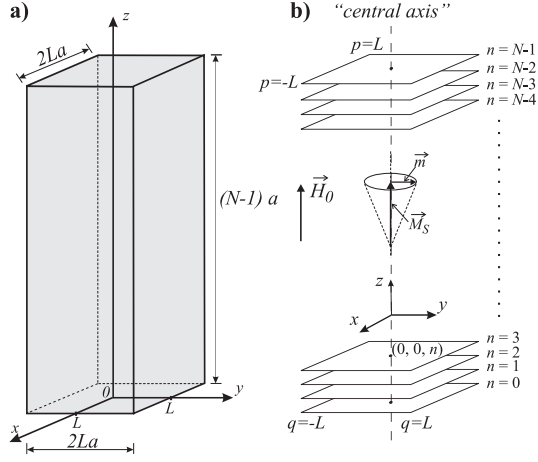}
\caption{ (a) The rod sample considered here; the applied field, $\vec{H}_{0}$, sets the magnetic moments in the direction perpendicular to the
rod base ({\em i.e.} along the $z$-axis). The rod 'thickness' is $(N-1)a$, and the square base side width is $2La$ ($a$ denoting the lattice
constant). The magnetostatic waves are assumed to propagate along the $z$-axis ({\em i.e.} in the direction of the applied field). (b) The
planar rod model used in our calculations.} \label{konfprost}
%\end{center}
\end{figure}

We shall consider a dipolar system consisting of magnetic moments $\mu_{\vec{r}}$ arranged regularly in sites $\vec{r}$ of a simple cubic
crystal lattice and coupled by the dipole-dipole interactions. The system is assumed to form a rectangular prism with square base (see Figure
\ref{konfprost}a). Let the prism base determine the $(x, y)$-plane of a Cartesian reference
 system, with the $z$-axis perpendicular to this plane. The reference point $(0, 0, 0)$ shall be
 placed in the {\em central} site of the prism bottom.

Let us calculate magnetic field $\vec{h}_{\vec{R}}$  "produced" by all the prism dipoles in a site indicated by internal vector $\vec{R}$.
According to the classical formula (obtained using the linear approximation [\onlinecite{landau,patton84}]), field $\vec{h}_{\vec{R}}$ can be
expressed as follows (in the SI units):
\begin{equation}
\label{eq5} \vec {h}_{\vec{R}} = \frac{1}{4 \pi} \sum\limits_{\vec{r} \neq \vec{R}}
 {\frac{3(\vec{r}-\vec{R}) \left(\vec{\mu}_{\vec{r}} \cdot (\vec{r} -\vec{R})
 \right)-\vec{\mu}_{\vec{r}} |\vec{r}-\vec{R}|^{2}}{|\vec{r}-\vec{R}|^{5}}},
\end{equation}
the above sum involving all the sites {\em except} the reference point ({\em i.e.} the site with position vector $\vec{r} \equiv \vec{R}$).
The lattice planes parallel to the prism base shall be numbered with index $n \in \langle 0, N-1\rangle$ (see Figure \ref{konfprost}b), and the
sites within each plane indexed with vector $\vec{r}_{||}=a[p\hat{i}+q\hat{j}]$, defined by integers $p,q \in \langle -L,L\rangle $. This means
that the position of the sites in which the magnetic moments are located, indicated by vector $\vec{r}$, shall be defined by a set of three
integers, $(p, q, n)$:
\begin{eqnarray}
\vec{r} \equiv  [\vec{r}_{||},a n] \equiv a[p,q,n], \;\;\; p,q \in \langle -L,+L\rangle \;\;\; \mbox{and}\;\; n \in\langle 0,N-1\rangle ,
\label{pq}
\end{eqnarray}
$a$ denoting the lattice constant. Thus, the considered prism contains $N(2L + 1)^{2}$ magnetic moments. Below we shall focus on the magnetic
field on the $z$-axis only, assuming its direction to be solely allowed for magnetic wave propagation. Hence, we put $\vec{R} \equiv a[0,0,n']$,
where $n' \in \langle 0,N-1\rangle $, and re-index the dipole field: $\vec{h}_{n'} \equiv \vec{h}_{\vec{R}}$. Additionally, we shall assume that
all the magnetic moments within {\em a single plane} $n$ are identical, {\em i.e.}: $ \vec{\mu}_{n} \equiv \vec{\mu}_{[p,q,n]}$, for any $p$ and
$q$. It is convenient to introduce here the notion of {\em magnetization}, a phenomenological quantity, which in the considered case of simple
cubic lattice can be defined as follows: $ \vec{M}_{n}=\vec{\mu}_{n}/a^{3}$. Then, (\ref{eq5})  becomes:
\begin{eqnarray}
\label{eq7a} \vec {h}_{n'} = \frac{1}{4\pi} \sum\limits_{n}D_{n,n'} \left[\hat{i}M_{n}^{x}+ \hat{j}M_{n}^{y}-2 \hat{k}M_{n}^{z} \right],
\end{eqnarray}
where we have introduced a matrix $\hat{D}$ whose elements, $D_{n,n'}$, are defined as follows:
\begin{eqnarray}
D_{n,n'} \equiv  \sum_{p,q}{}^\prime \frac{\frac{1}{2}\left(p^{2}+q^{2}\right) - (n-n')^{2}}{\left[ p^{2}+q^{2}+\left(n-n'\right) ^{2}\right]
^{\frac{5}{2}}}. \label{L}
\end{eqnarray}
It should be remembered that site $(0,0,n')$ is excluded from the sums appearing in the  equation (\ref{L}). Equation (\ref{L}), defining
elements $D_{n,n'}$, can be interpreted as the definition of a
 {\em dipolar} matrix $\hat{D}$ composed of these elements; as we shall see this
 matrix shall play an important role in
deducing properties of magnetic modes. As a consequence of (\ref{L}) matrix $\hat{D}$ is symmetric, {\em i.e.} $D_{n,n'} \equiv D_{n',n}$. Let
us introduce a new variable, defined as follows: $\delta= n - n'$; $\delta$ measures the distance between planes $n$ and $n'$. Consequently, we
can write:
\begin{eqnarray}
D_{n,n'} \equiv  D_{\delta} = D_{n',n} \equiv D_{-\delta}; \;\;\;\;\;\;\; D_{\delta} = \sum_{p,q}{}^\prime
\frac{\frac{1}{2}\left(p^{2}+q^{2}\right) - \delta^{2}}{\left[ p^{2}+q^{2}+\delta^{2}\right] ^{\frac{5}{2}}}. \label{DD}
\end{eqnarray}

Note that the considered system consists of a finite number of planes (index $n$ taking values $n = 0,1,2,\cdots,N-1$), so the set of values
available to $\delta$ depends on the
 reference plane, $n'$, with respect to which the distance is measured. However, the following
 condition must always be satisfied:
\begin{eqnarray}
0\leq n'+\delta \leq N-1. \label{DD2}
\end{eqnarray}
A sum over {\em all}  the system planes shall be in use below; this sum shall be denoted as: $\sum_{\delta} {}^{n'}$, superscript $n'$
indicating that the summing is performed on the planes neighbouring with $n'$, including plane $n'$. Therefore, $\delta$ takes the following
values: $ \delta = 0, \pm 1, \pm 2\, ...$ its lower and upper limits being determined by condition (\ref{DD2}).

Up to now, the direction of dipole arrangement has not had much importance in our reasoning. Now we shall consider the case with dipoles
arranged along the $z$-axis only.

\section{ Dynamics of the dipolar system}

 In this paragraph we shall consider a magnetic prism placed in a static magnetic field, $H_{0}$, applied along the $z$-axis (Figure
\ref{konfprost}). Field $H_{0}$ is assumed to be strong enough to arrange {\em all} the magnetic moments along the $z$-axis. Then, the
magnetization vector can be regarded as a superposition of two components: static (parallel to the $z$-axis) and dynamic (lying in the
$(x,y)$-plane):
\begin{eqnarray}
\vec{M}_{\vec{R}}=M_{S} \hat{k} + \vec{m}_{\vec{R}};   \label{Ms}
\end{eqnarray}
$M_{S}$  is the static magnetization, assumed to be homogeneous throughout the sample, and vector $\vec{m}$  denotes the dynamic magnetization,
perpendicular to $\vec{M_{S}}$. Similarly, the dipole field, $\vec{h}_{n'}$, can be resolved into two components: static, $\vec{h}_{n'}^{s}$
(parallel to the $z$-axis), and dynamic, $\vec{h}_{n'}^{d}$ (lying in the $(x,y)$-plane):
\begin{eqnarray}
\vec{h}_{n'} =\vec{h}_{n'}^{s}+\vec{h}_{n'}^{d}. \label{statdyn}
\end{eqnarray}
These two components of the dipole field can be easily found from (\ref{eq7a}). By replacing the third component of the magnetization vector
with the static magnetization ({\em i.e.} by putting $M_{n}^{z} \equiv M_{S}$), and the two other components, $M_{n}^{x}$ and $M_{n}^{y}$,
 with the respective components of the dynamic magnetization, $m_{n}^{x}$  and $m_{n}^{y}$,
 the following formulae are obtained:
\begin{eqnarray}
\vec{h}_{n'}^{s}  = -\left[ \frac{1}{2 \pi} \sum_{n} D_{n,n'}\right] M_{S} \hat{k}, \;\;\;\;\; \vec{h}_{n'}^{d} =  \frac{1}{4 \pi} \sum_{n}
D_{n,n'} \vec{m}_{n}, \label{hstat1}
%\label{hdyn1}
\end{eqnarray}
element $D_{n,n'}$ being defined by (\ref{L}).

The magnetic moment dynamics is described by the phenomenological Landau-Lifshitz equation (LL):
\begin{eqnarray}
\frac{\partial \vec{M}_{\vec{R}}}{\partial t} =\gamma \mu_{0} \vec{M}_{\vec{R}} \times \vec{H}_{eff,\vec{R}}, \label{LL1}
\end{eqnarray}
$\vec{H}_{eff,\vec{R}}$  denoting the effective magnetic field acting on the magnetic moment in site $\vec{R}$. This effective field is a
superposition of two terms only: the applied field, $\vec{H}_{0}$, and the field $\vec{h}_{n'}$
 produced by the magnetic dipole system:
\begin{eqnarray}
\vec{H}_{eff,\vec{R}} \equiv \vec{H}_{eff,n'}=\vec{H}_{0} + \vec{h}_{n'} = \left( H_{0}+h_{n'}^{s}\right) \hat{k}+ \vec{h}_{n'}^{d};
\end{eqnarray}
the above-introduced dipole field components (static and dynamic) being defined by (\ref{hstat1}).

The LL equation becomes:
\begin{eqnarray}
\frac{\partial \vec{m}_{n'}}{\partial t} =\gamma \mu_{0} \left(M_{S}\hat{k}+\vec{m}_{n'} \right) \times \left(\left( H_{0}+h_{n'}^{s}\right)
\hat{k}+ \vec{h}_{n'}^{d}\right).
 \label{LL2}
\end{eqnarray}
We shall solve it using the linear approximation, {\em i.e.} neglecting all the terms with $\vec{m}$ squared. Assuming the standard harmonic
time-dependence of the solutions: $\vec{m}_{n'} \sim e^{-i\omega t}$, (\ref{LL2}) becomes:
\begin{eqnarray}
-i\omega \vec{m}_{n'} =\gamma \mu_{0}\left[ M_{S}\hat{k} \times \vec{h}_{n'}^{d} + \vec{m}_{n'}  \times \left( H_{0}+h_{n'}^{s}\right) \hat{k}
\right]
 \label{LL3}
\end{eqnarray}
or, using the properties of vector product:
\begin{eqnarray}
-i\omega \vec{m}_{n'} =\gamma \mu_{0} \hat{k} \times\left[ M_{S} \vec{h}_{n'}^{d} - \vec{m}_{n'}  \left( H_{0}+h_{n'}^{s}\right) \right].
 \label{LL4}
\end{eqnarray}
Through replacing the dipole field dynamic and static components with their explicit expressions (\ref{hstat1}) we obtain:
\begin{eqnarray}
-i\omega \vec{m}_{n'} =\gamma \mu_{0} \hat{k} \times\left[ M_{S} \frac{1}{4 \pi} \sum_{n} D_{n,n'} \vec{m}_{n} -\vec{m}_{n'} \left(
H_{0}-\frac{M_{S}}{2 \pi} \sum_{n} D_{n,n'} \right) \right],
 \label{LL5}
\end{eqnarray}
or, after bilateral multiplication by $-4\pi(\gamma \mu_{0} M_{S})^{-1}$:
\begin{eqnarray}
i \Omega \vec{m}_{n'} = \hat{k} \times\left[ \vec{m}_{n'}  \left( \Omega_{H} -2 \sum_{n} D_{n,n'}\right) -\sum_{n} D_{n,n'} \vec{m}_{n} \right],
 \label{LL6}
\end{eqnarray}
$\Omega$  and $\Omega_{H}$ denoting the {\em reduced frequency} and the {\em reduced field}, respectively, defined as follows:
\begin{eqnarray}
\Omega \equiv \frac{4 \pi \omega}{\gamma \mu_{0} M_{S}} \mbox{\hspace{0.5cm} and \hspace{0.5cm}} \Omega_{H} \equiv \frac{4 \pi H_{0}}{M_{S}}.
\label{omega}
\end{eqnarray}
Two complex variables are now introduced for convenience: $m^{\pm}_{n} = m_{n}^{x} \pm i m_{n}^{y}$; with these new variables, (\ref{LL6})
splits into two independent {\em identical} scalar equations for $m^{+}_{n}$ and $m^{-}_{n}$; this means we are dealing with magnetostatic waves
polarized {\em circularly}. Therefore, it is enough to consider only one of these two equations, {\em e.g.}  that for $m^{+}_{n}$:
\begin{eqnarray}
\Omega m^{+}_{n'} = m^{+}_{n'} \left( \Omega_{H} -2 \sum_{n} D_{n,n'}\right) -\sum_{n} D_{n,n'} m^{+}_{n}.
 \label{LL7}
\end{eqnarray}
The above equation can be rewritten as follows:
\begin{eqnarray}
\Omega m^{+}_{n'} = m^{+}_{n'}  H(n') -\sum_{\delta=1,2...} D_{\delta} m^{+}_{n' \pm \delta},
 \label{LL7c}
\end{eqnarray}
where we introduced the following abbreviation denoting {\em the local field}:
\begin{eqnarray}
H(n') \equiv \Omega_{H} -D_{0}- 2 \sum_{\delta=0}{}^{n'} D_{\delta} \equiv \Omega_{H}+\Omega^{d}+\Omega_{n'}^{s}, \label{omegan}
\end{eqnarray}
with $\Omega^{d} \equiv - D_{0}$ and $\Omega_{n'}^{s} \equiv -2\sum_{\delta=0}{}^{n'} D_{\delta}$ meaning the ({\em reduced}) contributions to
the local field coming, respectively, from the dynamical and statical parts of the demagnetizing field. Note that, the eigenvalues $\Omega$
(being reduced frequencies) correspond to magnetostatic waves propagating in the direction of the applied field, {\em i.e. along the central
axis} shown in Figure  \ref{konfprost}.

In the remaining part of our work we will be considering only the rod-shaped samples, {\em i.e.} starting from this point we always assume $2L<
N-1$. We also assume particular values for $\mu_{0} H_{0}=0.2T$ and $M_{S}=0.139 \cdot 10^{6} Am^{-1}$ (YIG magnetization) with resulting value
for the reduced field $\Omega_{H}=14.374$. However, we have to emphasis that selection of this particular value for $\Omega_{H}$ is not
essential for results to be presented in subsequent sections of this work, since the distribution of eigenvalues $\Omega$ and profiles of modes
associated with them are not sensitive to the choice of particular $\Omega_{H}$ value: the particular  value of $\Omega_{H}$ only sets the {\em
whole} spectrum in a given frequency region and if $\Omega_{H}$ changes the whole spectrum is shifted to another region, but the {\em relative }
distribution of mode eigenfrequencies remains unchanged.

\section{Mode frequencies and amplitude profiles}

We shall investigate magnetostatic excitations in a rod of the plane size $2La \equiv 20a$ ($a$ being the lattice constant). The rod consists of
$N$ planes normal to the $z$-axis and numbered with index $n$, ranging from $n$=0 to $n=N-1$, as indicated in Figure \ref{konfprost}. The
effective dipole field calculation procedure applied in the previous paragraphs allows to find the field in $z$-axis points only, $i.e.$ along
the rod \textit{central axis} passing through opposite rod face centers; this is the idea of the approximation used throughout this study, and
henceforth referred to as \textit{central-axis approximation}.

The whole sample is assumed to be magnetized uniformly (the corresponding magnetization value being $M_{S})$ and in the direction of the
external field, applied along the z-axis. The $z$-axis direction shall be also the only one allowed for propagation of the magnetostatic waves
studied in this paper. With these assumptions, the problem of motion -- to be solved on the basis of equation (\ref{LL1}) -- reduces to a single
dimension in the space of variable $n$; the domain of the investigated motion is the interval $n\in (0, N-1)$, between two opposite rod face
centers.

\begin{figure}
%\begin{center}
\includegraphics[width=8cm ]{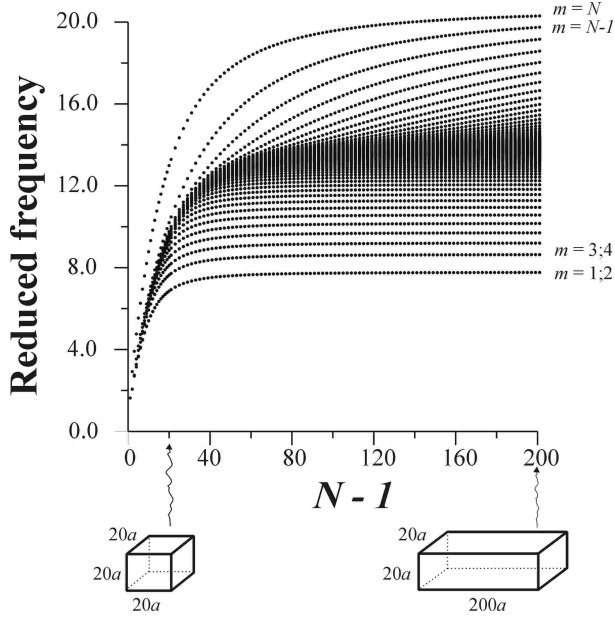}
\caption{The discretized magnetostatic mode frequencies \textit{vs.} the rod length, $(N-1)a$; $m$ indicates the mode number.}\label{fig2}
%\end{center}
\end{figure}

Figure  \ref{fig2} presents the discrete spectra of numerically calculated magnetostatic mode frequencies in a rod of variable length; the
spectrum evolution with increasing $N$ is visualized by the depicted frequency branches, each corresponding to one mode of a fixed number $m$.
The plot shows clearly that the frequencies stabilize after reaching some critical length; we note that lower modes stabilize at lower
"critical" values of length $N$ than higher modes.

\begin{figure}
\includegraphics[width=7cm ]{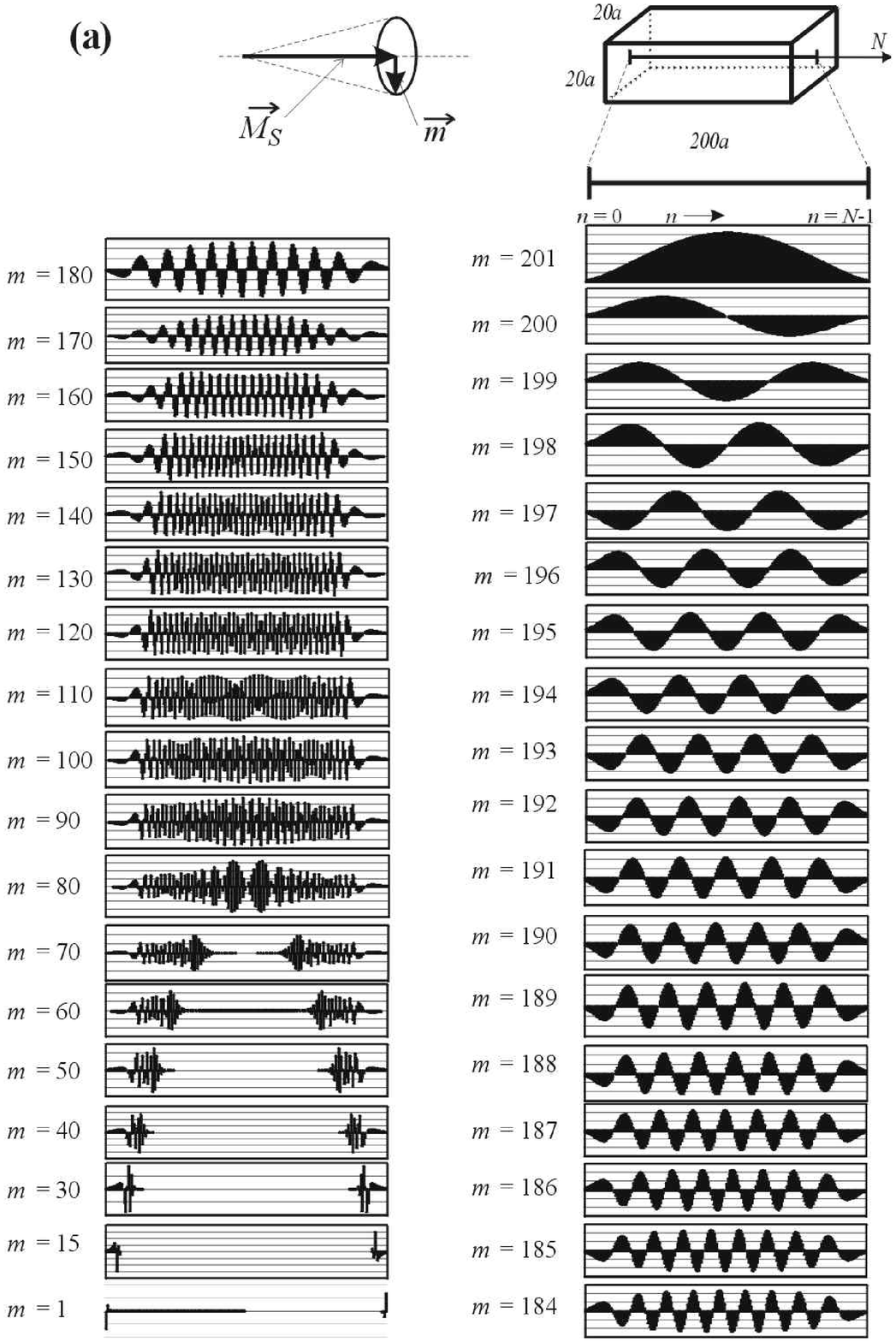}
\includegraphics[width=7cm ]{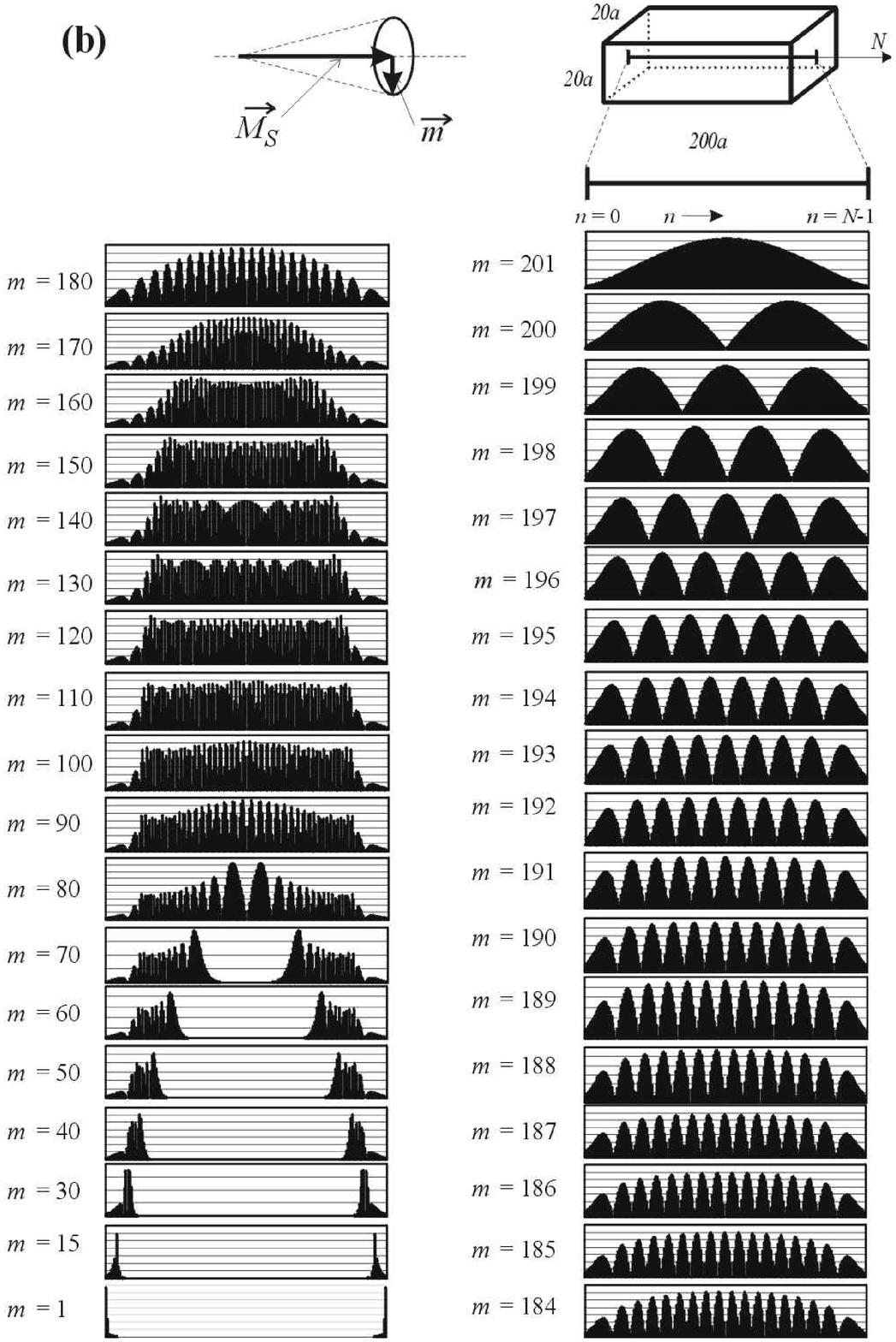}
\caption{Numerically calculated magnetostatic mode profiles in a rod of length 200$a$ depicted along the central axis (indicated in Figure
\ref{konfprost}) showing separately: (a) the dynamical magnetization $m^{+}$ relative values, and (b) the corresponding absolute values, $\vert
m^{+}\vert $.}\label{fig3}
\end{figure}

In order to discuss in detail the properties of magnetostatic mode spectrum in rods, let's consider a sample with dimensions $20a \times 20a
\times 200a$, {\em i.e.} a square rod with length exceeding ten times the base side. The plot shown in Fig. \ref{fig2} indicates that the mode
spectrum corresponding to this aspect ratio is stable enough to assume that no significant changes would result from further rod elongation.
Fig. \ref{fig3}a shows eigenmode profiles numerically calculated in the considered rod. Though showing only selected modes, the depicted mode
profile sequence provides full information on amplitude variation with mode frequency. The features of the mode profiles shown in Fig.
\ref{fig3}a allow to classify all the modes obtained into four qualitatively distinct groups: (a) slow-oscillating bulk modes, comprising {\em
ca.} 20 modes with highest frequency values; the number of oscillations grows as mode frequency \emph{decreases}; (b) modes $m=170$ to $m=90$,
with different character of amplitude profile oscillations in two distinct regions: a central region with fast-oscillating amplitude, and outer
regions with slow-oscillating amplitude; (c) modes $m=3$ to $m=80$, whose amplitudes are \emph{completely} suppressed in a central region,
communicating on each side with a fast-oscillation zone followed by a slow-oscillation region; (d) two surface-localized modes, $m=1$ and $m=2$,
with amplitudes concentrated almost exclusively on the sample surface. For the purposes of further analysis, the mode amplitude \emph{absolute
value} profiles are also depicted in Fig. \ref{fig3}b. The mode profiles shown in Fig. \ref{fig3}a,b allow to identify the modes in groups (a)
and (d) as \emph{backward bulk extended modes} and \emph{surface localized modes}, respectively; both these mode types are well-known from
literature. The remaining two groups, to our best knowledge not yet reported in literature, represent a new quality, and therefore need to be
discussed in detail. Modes from group (b) - henceforth referred to as \emph{comb modes}, because of the characteristic 'comb' profile of their
amplitudes in a central region - are characterized by two wave numbers: high in the central region, low in the outer regions. A characteristic
feature of modes from group (c) is complete amplitude vanishing in a central 'zero region'; since this zero region broadens as mode frequency
decreases, the non-zero amplitudes in the adjacent outer regions are thinning down as they descend.  For this reason, modes from group (c) will
be henceforth referred to as \emph{bulk-dead modes}. As we shall see further into this paper, comb and bulk-dead modes appear as a result of rod
elongation along a distinguishing axis.

\section{Internal dipolar field nonhomogeneity {\em vs.} comb and bulk-dead modes}

\begin{figure}
\includegraphics[width=8cm ]{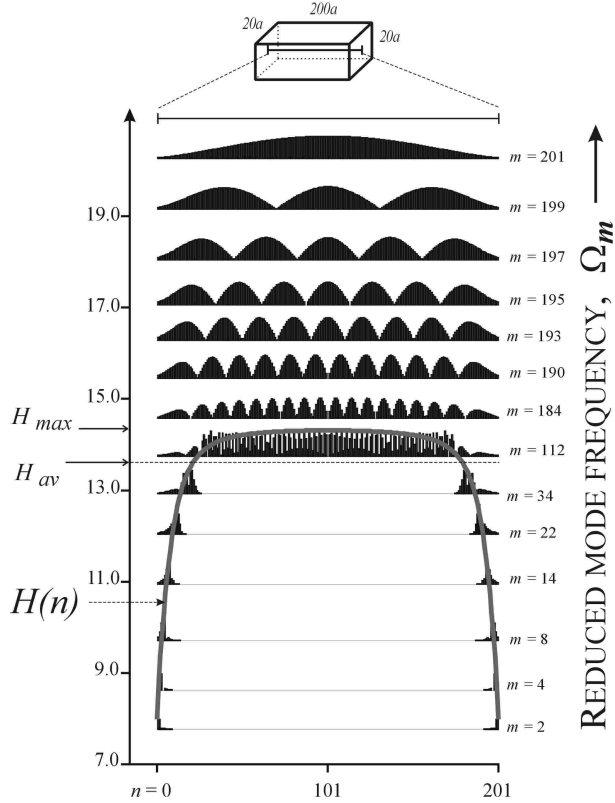}
\caption{Magnetostatic modes in a rod of length 200$a$ confronted with the spatial distribution of the local field $H(n)$ (bold line) depicted
along the rod central axis: only absolute values $|m^{+}|$ mode amplitudes are shown here. A striking feature is that the local field line
borders the "empty" region (where mode amplitudes are suppressed) of \emph{bulk-dead modes}; the \emph{bulk-extended} mode frequencies lie
\textit{above} the local field range ($\Omega_{m}> H_{max}$).}\label{fig4}
\end{figure}

\begin{figure}[htp]
\includegraphics[width=9cm ]{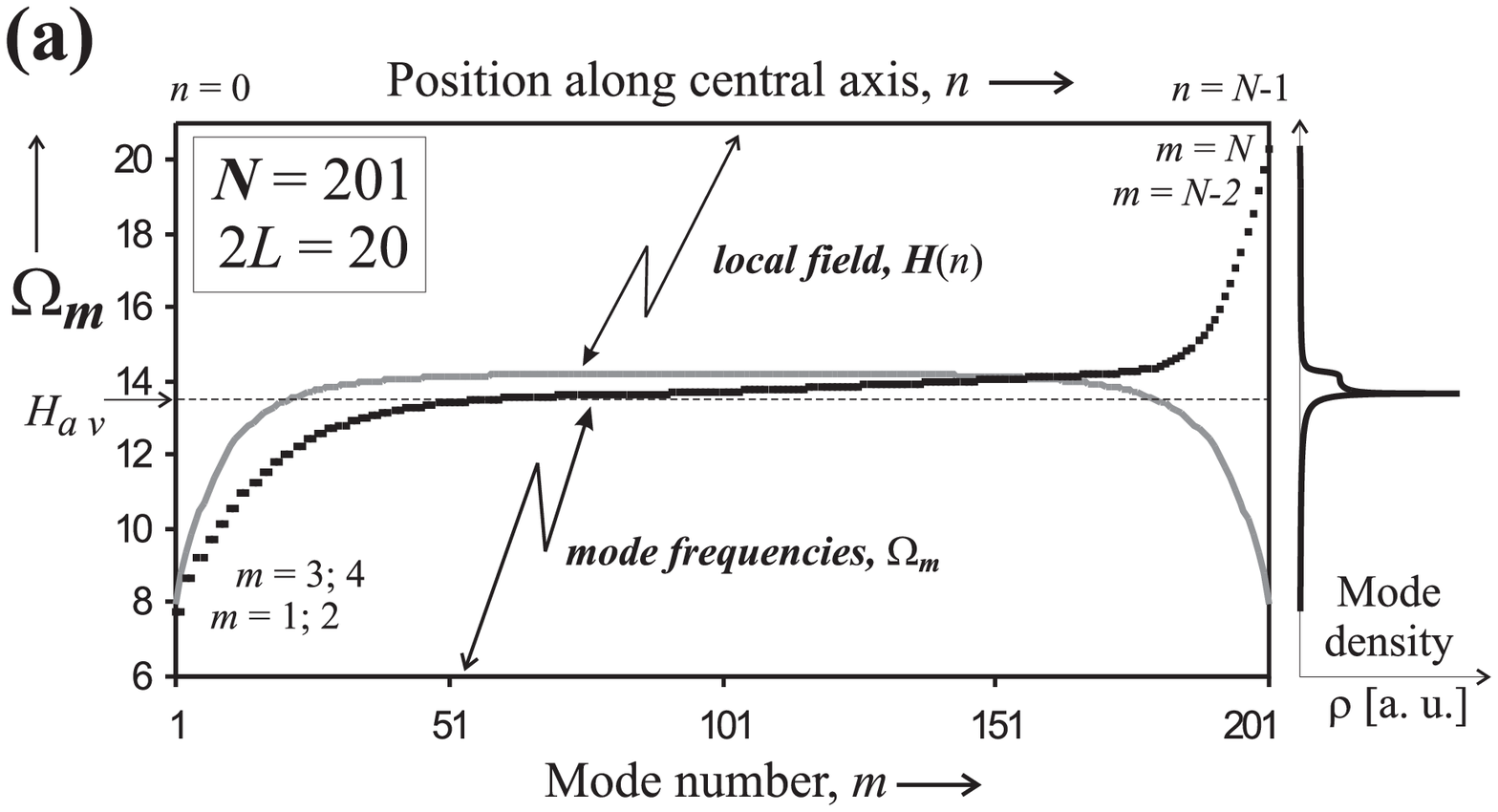}
\hfill
\includegraphics[width=6cm ]{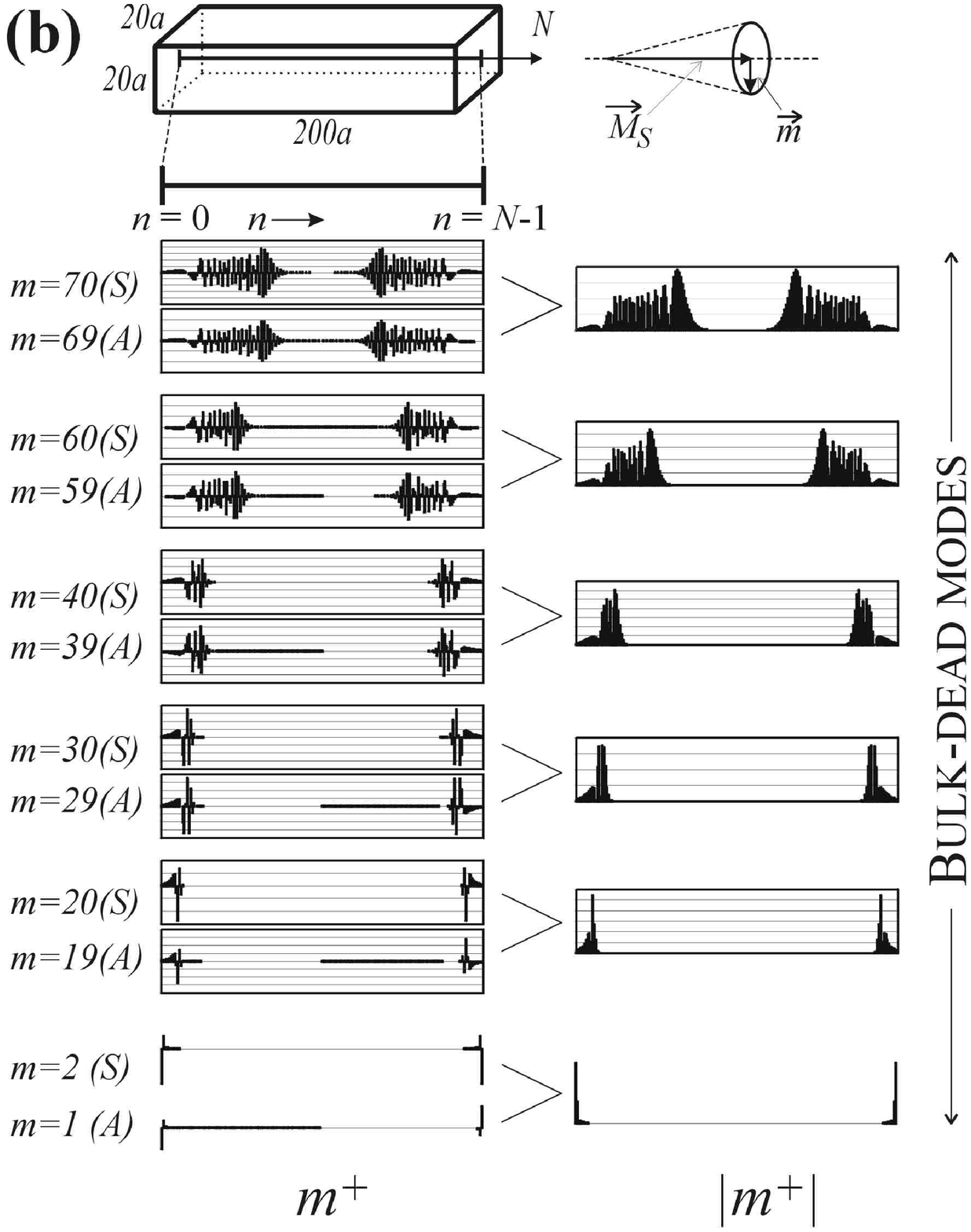}
\caption{(a) The discretized magnetostatic mode frequencies (black squares) \textit{vs.} the mode number, $m$. The bold line represents the
spatial distribution of the local  field $H(n)$  along the central axis; the rod length is $200a$. (b) Note the double degeneration of the
lower-spectrum  (\emph{bulk-dead}) modes with $m=1-70$ ($S/A$ means, respectively, symmetric-antisymmetric mode).}\label{fig5}
\end{figure}

Now let's superimpose the calculated magnetostatic mode frequency spectrum and the profile of local field $H(n)$ plotted along the rod axis
(Fig. \ref{fig4}). A strong spatial inhomogeneity of the local field is apparent from the calculated field values; two regions, qualitatively
distinct in terms of inhomogeneity, can be distinguished: outer regions, with steep profile bias indicating high local field gradient, and a
central region, in which the profile bias is relatively mild. Note also that the average value of local field, $H_{av}$, can be regarded as the
border line between these two regions. The mode profiles, ordered by growing frequency, are superimposed on the local field curve, with energy
scale (on the vertical axis) maintained ({\em i.e.} common for the mode frequencies $H_{m}$ and the local field profile $H(n)$). An interesting
correlation occurs between the mode type and the position of the mode frequency with respect to the local field profile. All the modes with
frequency values above the local field maximum ($H_{max}$) are bulk modes. Those with frequency values between $H_{max}$ and  $H_{av}$ are comb
modes, and those with frequency values below $H_{av}$ are bulk-dead modes. Note that the fast-oscillation segment in comb mode profiles
coincides (approximately) with the low-gradient segment in the local field profile; at the same time,  the borders of the zero region in
bulk-dead mode profiles follow the local field profile in its high-gradient (outer) parts. Also, note a mode density peak in the immediate
vicinity of the average local field value $H_{av}$;  as much as 150 mode frequencies, representing approximately 3/4 of their total number, are
concentrated around this point! Further interesting conclusions can be drawn from superimposition of the complete mode frequency spectrum on the
local field profile (Fig. \ref{fig5}a). The first thing to be observed is a \emph{double degeneration} of bulk-dead modes ({\em i.e.} the modes
below $H_{av}$); the meaning of this degeneration of mode \emph{energy levels} is explained by Fig. \ref{fig5}b, showing profiles of all
bulk-dead modes with their spatial symmetry specified. Clearly, the degeneration of bulk-dead mode energy levels is associated with that of the
corresponding mode \emph{amplitudes}: in each degenerated pair, one state is symmetric and the other antisymmetric, both showing (see Fig.
\ref{fig5}b) identical \emph{amplitude absolute value profiles}. Mode degeneration disappears in the comb mode frequency range (which is
understandable, as symmetric states in this group have non-zero amplitudes in the rod \emph{center}, and thus amplitude degeneration is
impossible). Returning to Fig. \ref{fig5}a, an interesting thing to note is the linear character of comb mode frequency variation with mode
number $m$; the range of comb mode dispersion linearity falls in the immediate vicinity of $H_{av}$, the value at which - as we have already
anticipated on the basis of Fig. \ref{fig4} - mode density peaks (see the supplementary plot attached to the main one on its right side).

\section{ Magnetostatic mode resonance spectrum and outlooks}

\begin{figure}
\includegraphics[width=15cm ]{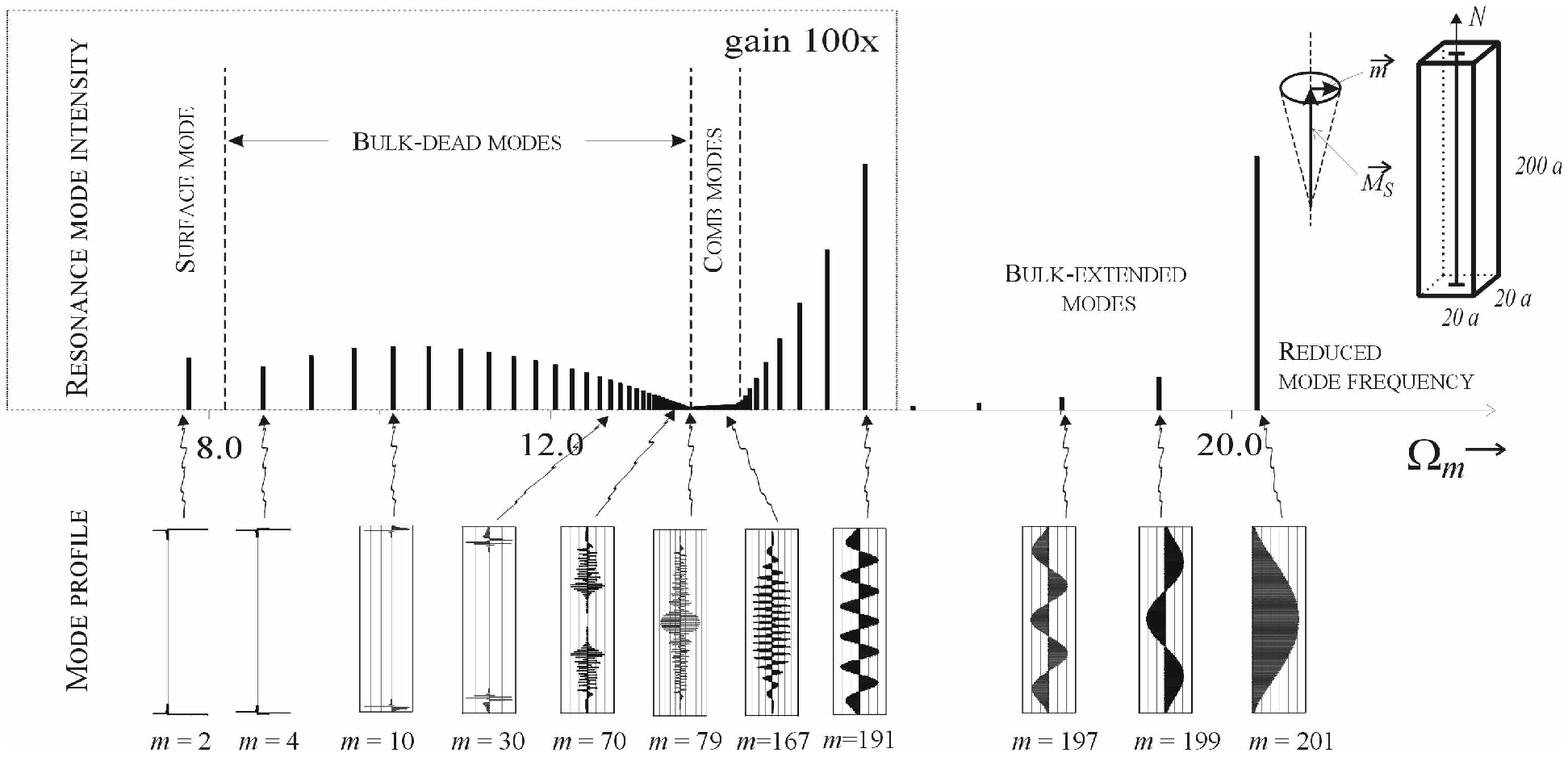}
\caption{Resonance spectrum of magnetostatic modes in the rod of the length $200a$ and the aspect ratio 10. Note the existence of the
nonabsorbing frequency region corresponding to comb modes.}\label{fig6}
\end{figure}

 The presence of comb and bulk-dead modes in a rod spectrum results in a very interesting resonance effect, illustrated in Fig. \ref{fig6}.
 The top part shows the SWR spectrum calculated for a $20a \times 20a \times 200a$ rod; the modes corresponding to the resonance lines are identified,
  through their profiles, in the bottom part of Fig. \ref{fig6}. The most intensive line corresponds to bulk mode $m=201$, having the highest
  frequency value. The intensity of subsequent resonance lines decreases systematically with mode frequency, as can be expected
  from the growing number of bulk-extended mode nodes. The resonance intensity nearly vanishes in the comb mode frequency region,
  to revive again and \emph{increase} in the bulk-dead mode frequency region. Such resonance absorption minimum \emph{inside} a spectrum is very
  uncommon [\onlinecite{puszkarski79}], and its existence may be of practical importance: due to the peak mode density in its vicinity ({\em cf.} Fig. \ref{fig5}a), such a minimum
  would indicate a frequency range in which substantial reduction of magnetostatic noise could be expected .

Is there any evidence of the comb mode existence reported in the literature? We believe such an evidence to be provided {\em e.g.} by Bayer {\em
et al.} [\onlinecite{bayer04}], in a study based on time-resolved Kerr microscopy, used to detect collective spin wave modes in inhomogeneously
magnetized $Ni_{0.81}Fe_{0.19}$ thin-film stripes. The permalloy stripes investigated were 18 $nm$ thick, 2.3 $\mu m$ wide and 1 $mm$ long, and
the experiment was performed with an in-plane magnetic field oriented along the stripe short axis. A new type of excitation is distinguished by
the authors among the spin wave modes observed, and named "crossover modes": though spanning the entire stripe, these "new" modes are only
detectable at the edges. To explain this feature, the authors refer to two different spatial regimes: a small wave-vector region near the edge
of the stripe, and a large wave-vector region surrounding the stripe center. Although the geometry of the samples used in the experiments by
Bayer {\em et al.} (stripe) is different than that of the samples considered in this paper (rod), the properties of the experimentally found
"crossover modes" strikingly coincide with those of the "comb modes" described here. Moreover, both mode types (crossover and comb) have the
same origin, appearing as a result of spatial inhomogeneity of the local field along the direction of mode formation. The only difference lies
in the origins of this inhomogeneity: the spatial inhomogeneity of the effective field in the samples used by Bayer \emph{et al.} is due to
inhomogeneous magnetization distribution across the stripe, while in our study, the effective field inhomogeneity is a consequence of a spatial
inhomogeneity of the respective demagnetization coefficient (the sample magnetization is assumed to be homogeneous).

Let's get back once more to the resonance spectrum resulting from our calculations and depicted in Fig. 6. Note that this spectrum contains a
very peculiar segment, corresponding to excitation of bulk-dead modes. The peculiarity of this segment lies in the fact that the resonance line
intensities vary \emph{non-monotonically} with increasing energy: a rising part and a falling one can be distinguished, with a local maximum
between them, located in the central part of the spectrum. This type of behaviour is very unusual, as resonance intensities in multi-peak
ferromagnetic resonance spectra tend to be \emph{monotonic} functions of energy. Therefore, we searched the available literature for a similar
spectrum segment in the results of experimental resonance measurements (performed in rods), and we found an old paper by Patton and Schlomann
[\onlinecite{patton68}] reporting exactly the resonance spectrum segment we have been looking for. It appeared as a result of their study quite
unusually, and only due to the authors' very original idea to distinguish, in their resonance measurements, two spectrum components: one
associated with the central part of the rod, the other with its border  parts.

Patton and Schlomann investigated an axially magnetized cylindrical 0.075 inch (diameter) 3 inch (length) YIG rod; the central part in which a
separate resonance could be obtained was 0.4 inch long. The original measurement setup, conceived by the authors, allowed a resonance spectrum
to be measured first in the whole rod and then, separately, only in its central part. The spectrum resulting only from the absorption in the rod
borders could be deduced by differentiation from both spectra obtained (Fig. \ref{fig4}a in the quoted study [\onlinecite{patton68}]). A theory
of magnetostatic modes in axially magnetized rods, developed by Joseph and Schl\"{o}mann [\onlinecite{joseph61}], was used as a basis of
interpretation of the measurements, but only the resonance positions of the magnetostatic modes \emph{calculated} on this basis were identified.
The resonance intensities were not interpreted theoretically; the results obtained within the present study seem to allow to fill in this gap.
We shall focus on the series of resonance lines associated with the modes excited in the rod borders as shown in [\onlinecite{patton68}] in
Fig.4a. Strikingly, this resonance pattern is qualitatively identical with the segment of our spectrum (Fig. \ref{fig6}) identified as
associated with bulk-dead modes. Indeed, a deeper sense is hidden behind this apparent analogy: since bulk-dead modes contain a dead-amplitude
central zone, their amplitudes being significant only in the rod borders, obviously, their resonance excitation can come from the borders only,
exactly as indicated by the Patton and Schl\"{o}mann's experiment! Although our theory deals with a square rod, while the Patton and
Schl\"{o}mann's experiment was performed on a cylindrical one, we anticipate that this geometrical difference is not essential for the
above-discussed intensity patterns, since our in-plane sums (\ref{DD2}) are only slightly modified when a square is replaced by a circle
[\onlinecite{hp}]. Thus, anticipating bulk-dead modes on a theoretical basis in the present paper, we believe that  Patton and Schl\"{o}mann's
experiment [\onlinecite{patton68}] provides a proof of their existence.

\begin{acknowledgements}
This work was supported by the Polish Committee for Scientific Research through the projects KBN - 2P03B 120 23 and PBZ-KBN-044/P03-2001.
\end{acknowledgements}


\begin{thebibliography}{15}
\bibitem[*]{correspond} Corresponding author: H. Puszkarski, Surface
Physics Division, Institute of Physics, Adam Mickiewicz University, ul. Umultowska 85, 61-614 Poznan, Poland; \textit{Email address}:
henpusz@amu.edu.pl
\bibitem{givord03} D. Givord, Europhysics News, {\bf 34/6},
219 (2003).
\bibitem{2} M. Pardavi-Horvath, and J. Yan, IEEE Trans. Magn., {\bf 39}, 3154 (2003).
\bibitem{levy04} J.-C. S. L\'evy, D. Mercier, H. Puszkarski, and M. Krawczyk, Recent Res. Devel. Magnetism \& Magnetic Mat. {\bf 2}, 1 (2004).
\bibitem{puszkarski05} H. Puszkarski, M. Krawczyk, and J.-C. S. L\'evy, Phys. Rev. B {\bf 71}, 014421 (2005).
\bibitem{hp} H. Puszkarski, M. Krawczyk, and J.-C. S. L\'evy,  to be published.
\bibitem{landau} L.D. Landau,  and E.M. Lifshitz, The Classical Theory of
Fields (Pergamon Press, Oxford, 1975).
\bibitem{patton84} C.E. Patton, Physics Reports {\bf 103}, 251 (1984).
\bibitem{puszkarski79} H.  Puszkarski, Progress in Surface Science {\bf 9}, 191 (1979).
\bibitem{bayer04} C. Bayer, J.P. Park, H. Wang, M. Yan, C.E. Campbell, and P.A. Crowell, Phys. Rev. B {\bf 69}, 134401 (2004).
\bibitem{patton68} C.E. Patton, and E. Schl\"{o}mann, IEEE Transactions on Magnetics {\bf MAG-4}, 596 (1968).
\bibitem{joseph61} R.I. Joseph and E. Schl\"{o}mann, J. Appl. Phys. {\bf 32}, 1001 (1961).




\end{thebibliography}
\end{document}